\documentclass{article}

\usepackage{PRIMEarxiv}

\usepackage[utf8]{inputenc} 
\usepackage[T1]{fontenc}    
\usepackage{hyperref}       
\usepackage{url}            
\usepackage{booktabs}       
\usepackage{amsfonts}       
\usepackage{nicefrac}       
\usepackage{microtype}      
\usepackage{lipsum}
\usepackage{fancyhdr}       
\usepackage{graphicx}       
\graphicspath{{media/}}     
\usepackage{amsmath}  
\usepackage{pifont}  
\title{BOgen: Generating Part-Level 3D Designs Based on User Intention Inference through Bayesian Optimization and Variational Autoencoder

}

\author{
  Seung Won Lee\\
  Department of Interior Architecture Design \\
  Hanyang University\\
  Seoul \\
  \texttt{lswgood0901@gmail.com} \\
   \And
  Jiin Choi  \\
  Department of Interior Architecture Design \\
  Hanyang University\\
  Seoul \\
  \texttt{jiin4900@gmail.com} \\
   \And
  Kyung Hoon Hyun\thanks{Corresponding author. E-mail address: hoonhello@hanyang.ac.kr (K. H. Hyun).}  \\
  Department of Interior Architecture Design \\
  Hanyang University\\
  Seoul \\
  \texttt{hoonhello@hanyang.ac.kr} \\
}

\begin{document}
\maketitle

\begin{abstract}
Advancements in generative artificial intelligence (AI) have introduced various AI models capable of producing
impressive visual design outputs. However, when it comes to AI models in the design process, prioritizing outputs
that align with designers’ needs over mere visual craftsmanship becomes even more crucial. Furthermore,
designers often intricately combine parts of various designs to create novel designs. The ability to generate designs
that align with the designers’ intentions at the part level is pivotal for assisting designers. Hence, we introduced
BOgen, which empowers designers to proactively generate and explore part-level designs through Bayesian
optimization and variational autoencoders, thereby enhancing their overall user experience. We assessed BOgen’s
performance using a study involving 30 designers. The results revealed that, compared to the baseline, BOgen
fulfilled the designer requirements for part recommendations and design exploration space guidance. BOgen
assists designers in navigation and development, offering valuable design suggestions and fosters proactive design
exploration and creation.
\end{abstract}


\section{Introduction}

\begin{figure}[ht]
  \centering
  \includegraphics[width=\linewidth]{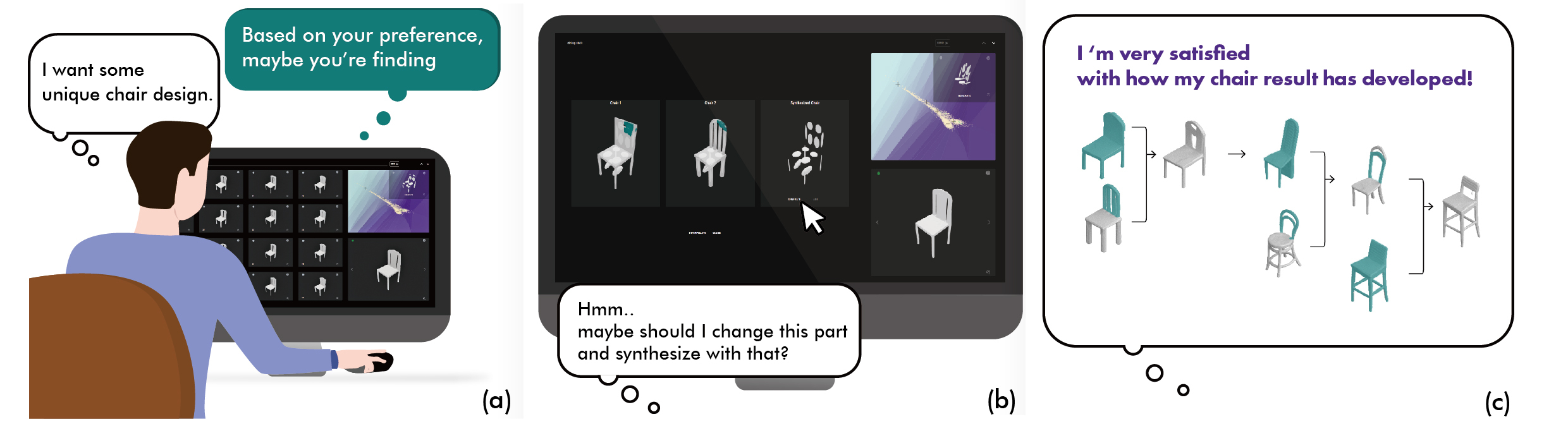}
  \caption{BOgen Overview: (a) Bogen recommends user-desired chairs by inferring user’s preferences; (b) BOgen
allows users to generate and synthesize part-level 3D designs; (c) BOgen guides users to navigate through vast
design space for design idea refinement.}
  \label{fig:1}
\end{figure}

Recently, there has been a surge in interest in Generative AI due to its capability to produce high-quality content with remarkable fidelity \cite{ramesh2022hierarchical,rombach2022high,von2023fabric,zheng2023locally}. However, 3D generative AI differs from 2D generative AI in having fewer paired text-shape data and limitations in terms of style and diversity \cite{liu20233dall,sanghi2023clip}. This presents many challenges and limitations in creating 3D designs that satisfy user requirements through 3D generative AI. Nonetheless, there have been proposals for 3D generative AI models that show impressive performance in 3D generation, including capabilities such as part-level inversion, generation, and interpolation \cite{hertz2022spaghetti,hui2022neural,koo2023salad}. For instance, Hertz et al. \cite{hertz2022spaghetti} introduced two types of latent concepts that represent the overall shape and fine details of 3D objects. Their model allows shape adjustments at the part-element level. Based on this, Koo et al. \cite{koo2023salad} developed a diffusion-based method for learning these latent variables and introduced a model for text-guided 3D generation. With these echnological advancements, labor-intensive and time-consuming 3D modeling tasks are now more accessible to general users and can be performed using simple text-based prompting. Furthermore, these 3D generative AI enables designers to explore various rich and highly informative design variations. However, when it comes to AI models in the design process, prioritizing outputs that align with designers’ needs over mere visual craftsmanship becomes even more crucial. Furthermore, designers often intricately combine parts of various designs to create novel designs. The ability to generate designs that align with the designers’ intentions at the part level is pivotal for assisting designers. 

Therefore, there is a need for methods that support designers iteratively \cite{son2022designer,son2022creativesearch}, enabling them to efficiently navigate a vast design space with limited time and computational costs. This has prompted numerous studies on generative design and AI, supporting designers in efficient design exploration \cite{pandey2023juxtaform,averkiou2014shapesynth,matejka2018dream}. These systems utilize interactive genetic algorithms and design optimization to efficiently navigate extensive design spaces and generate optimized designs based on user inputs. Additionally, several studies have developed systems using generative adversarial network (GAN) models to create and explore desired images through various user interfaces \cite{evirgen2022ganzilla,evirgen2023ganravel,dang2022ganslider}. However, despite numerous studies proposing user-friendly methods for navigating extensive design spaces, research is still limited in identifying specific designs desired by designers within these spaces,understanding part-level design intentions, and providing recommendations and guidance accordingly. Consequently, they are unable to address specific design problems and satisfy human psychological intentions because of failure to capture the designer's intention \cite{wu2023human}. Guiding designers to navigate a high-dimensional design space in real time based on their preferences is a formidable challenge. To address this issue, innovative methods that facilitate the exploration of a high-dimensional design space and estimate user intentions within that design space are required. To estimate design intentions, systems have been proposed that probabilistically interpret or optimize user behavior based on methodologies such as the Bayesian information gain framework or Bayesian optimization, providing appropriate feedback \cite{kadner2021adaptifont,koyama2022bo,lee2023bigaze,liu2017bignav,son2022bigexplore}. In addition, a method was proposed to map highdimensional design parameters onto a 2D latent space using variational autoencoders (VAE) \cite{danhaive2021design}. However, comprehensive research on the integrated methods and interfaces for exploring 3D generative AI design spaces remains scarce.

Building on prior research, this study aims to support design generation by inferring user preferences from the selection of design parts in a reduced design space termed the, "exploration map." We introduce BOgen, a collaborative human-AI system designed to generate 3D models by leveraging user behavior and an exploration map, especially in scenarios that demand  art-level generation and synthesis through 3D generative AI (Figure \ref{fig:1}; Video Supplementary Material). BOgen identifies and delivers design outcomes containing desired parts through part-level selection. It also presents a method for creating a design exploration map that facilitates real-time interaction within the time and computational constraints. Our approach involved six steps: 1) collecting 58,750 SALAD latent data to create the exploration map; 2) developing a VAE-based 2D exploration map; 3) developing preferential Bayesian optimization (PBO) for BOgen; 4) Integrating PBO, VAE, and SALAD; 5) developing a BOgen and UIonly interface; and 6) conducting a comparative experiment with 30 designers.

\section{Related Work}
\label{sec:headings}
\subsection{3D Generative AI in Design Exploration}
Recent advances in 3D generative AI models have demonstrated remarkable capabilities for producing high-quality results \cite{achlioptas2018learning,hertz2022spaghetti,hui2022neural,koo2023salad,zheng2023locally}. These models enable the generation, manipulation, and interpolation of 3D objects. For instance, Achlioptas et al. \cite{achlioptas2018learning} developed a model that generates point clouds from 3D GANs. Hui et al. \cite{hui2022neural} introduced a wavelet-based diffusion network capable of generating, manipulating, and interpolating highquality
3D objects at a part-level. Further advancements by Hao et al. \cite{hao2020dualsdf} and Hertz et al. \cite{hertz2022spaghetti} involved models that could control specific parts by decomposing the implicit shapes of objects, thereby facilitating part-level manipulations. Hertz et al.'s SPAGHETTI system notably enables part-level shape control through the use of extrinsic latent elements representing the overall shape and intrinsic latent elements representing fine details. Building on this, Koo et al. \cite{koo2023salad} refined output quality and segregated the training of low-dimensional extrinsic latents from high-dimensional intrinsic latents using a diffusion-based network. The i-th part of an object in SPAGHETTI and SALAD is represented by an extrinsic vector \( \mathbf{e}_i = \{c_i, \lambda_{i}^{1}, \lambda_{i}^{2}, \lambda_{i}^{3}, u_{i}^{1}, u_{i}^{2}, u_{i}^{3}, \pi_i\} \). Here, \( c \) represents the mean of the Gaussian  ixture, \( \lambda \) are the eigenvalues of covariance, \( u \) are the eigenvectors, and \( \pi \) represents the blending weights indicating the relative importance of each part. Hertz et al. and Koo et al. calculated these extrinsic vectors, that is, the approximate shape of a 3D Gaussian. However, the primary focus of these studies has been on perfecting the precision and quality of 3D generative outputs.

On the other hand, systems supporting design exploration through generative design based on interactive genetic algorithms and design optimization \cite{hyun2018balancing,ban20203d,khan2018generative}, as well as generative AI have been proposed \cite{evirgen2022ganzilla,zhou2021interactive}. Hyun and Lee \cite{hyun2018balancing} proposed a GA-based framework to support designers in making decisions based on styling strategies. Ban and Hyun \cite{ban20203d} suggested a framework that interpolates input sketches with those in a database to create design variations and, based on these variations, generates 3D models to support design exploration. Evirgen and Chen \cite{evirgen2022ganzilla} introduced GANzilla, a generative adversarial network (GAN)-based model that supports nonexpert users in creating desired images using scatter-and-gather techniques. Furthermore, Zhou et al. \cite{zhou2021interactive} proposed a model enabling users to intervene in the Bayesian optimization loop, adjusting the model's exploration and exploitation to create the desired melody compositions. However, there is a clear gap in research that aligns the power of generative AI with a designer’s unique intent across a myriad of design components. From a user-experience perspective, insights into how best to leverage these state-of-the-art 3D generative AIs are notably absent, especially when creating user-friendly interfaces to support the 3D generation process. While current leading studies emphasize the accuracy of the generation performance, their application in real-world design contexts remains largely unexplored. As pivotal as advancing AI performance is, devising ways for designers to utilize AI efficiently is crucial. Hence, this study seeks to bridge this gap by proposing methods and interfaces that integrate state-of-the-art 3D generative AIs into the design exploration process.

\subsection{User Intention Inference and Feedback Methods}
To generate objects that are meaningful and useful to designers, it is essential to enable design creation, combination, and exploration at a detailed element level that aligns with the designer’s intentions. However, whether such designs are useful can change depending on the design situation, and ultimately, designers should make that decision. To support this design process, design systems employing probabilistic models like Bayesian information gain (BIG) and Bayesian optimization (BO) have been proposed \cite{kadner2021adaptifont,koyama2022bo,lee2023bigaze,son2022bigexplore}. Son et al. \cite{son2022bigexplore} and Lee et al. \cite{lee2023bigaze} introduced systems that interpret user behavior probabilistically to provide design feedback based on information gain. Kadner et al. \cite{kadner2021adaptifont} proposed ’adaptifont,’ a BO-based font generation system that uses the user’s reading speed as an objective function. Koyama and Goto \cite{koyama2022bo} introduced preferential Bayesian optimization (PBO), which optimizes a function predicting the user’s desired goal based on their slider bar manipulation information. These systems track the design process by probabilistically interpreting user actions and offering design feedback that reflects the designer’s evolving intentions. However, as the information space increases, the computational cost of the BIG framework increases exponentially and BIG-based systems provide suboptimal design feedback to address this issue \cite{lee2023bigaze,liu2018bigfile,son2022bigexplore}. Similarly, BO operates best in lower-dimensional spaces, often requiring dimension reduction due to computational limitations \cite{kadner2021adaptifont}. In this context, Danhaive and Mueller \cite{danhaive2021design} proposed a system that learns design parameters and performance using a variational autoencoder and maps them onto a 2D space to generate designs. They introduced a graph that visualizes and navigates a complex highdimensional design space in lower dimensions. This allows users to visually grasp, compare and evaluate various design options. However, research on inferring the desired design elements from a reduced exploration map to support design creation and synthesis remains unexplored. Integrating the mentioned methods to infer user intentions, along with a streamlined exploration map approach, remains a significant challenge in proposing a 3D generative design system. Therefore, this study introduces a system that reduces the high-dimensional 3D generative latent to a more manageable exploration map, thereby guiding the design exploration process based on user preference estimations.

\section{BOgen}
\subsection{Conceptual Overview}
\begin{figure}[ht]
  \centering
  \includegraphics[width=\linewidth]{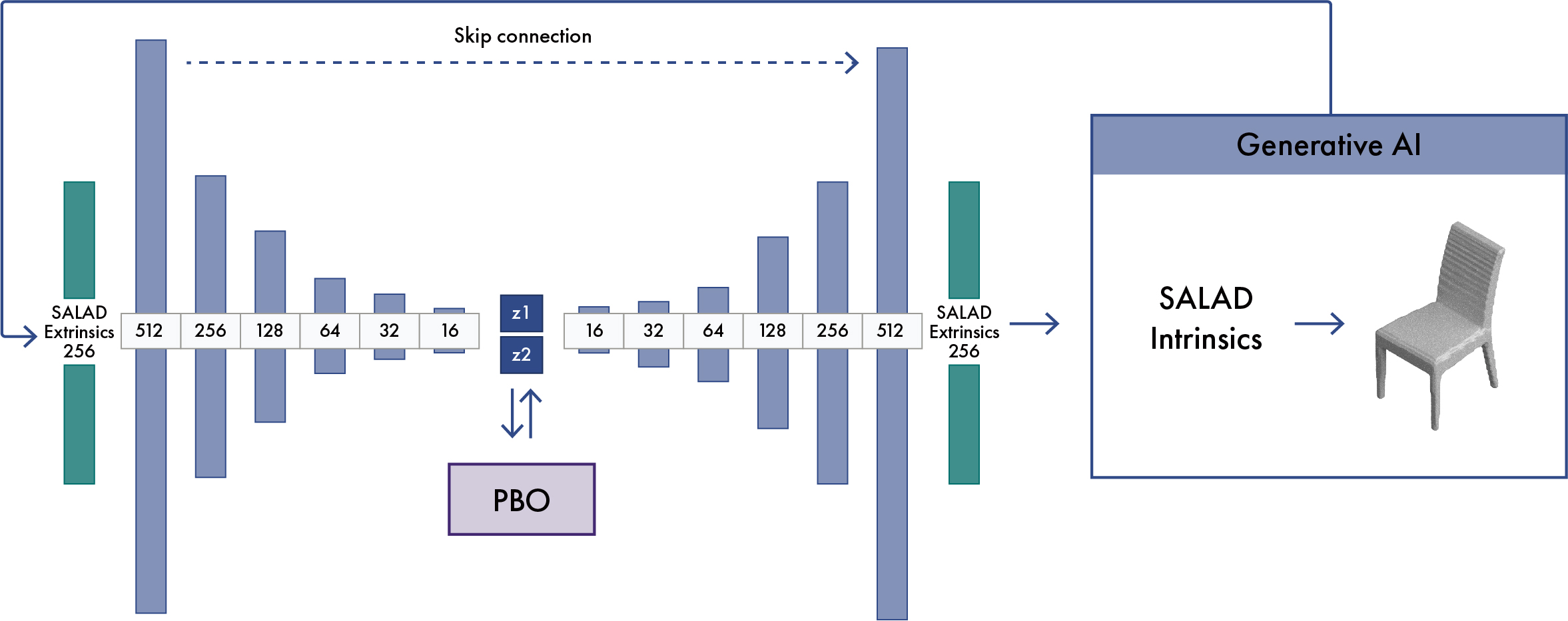}
  \caption{Bogen conceptual overview.}
  \label{fig:2}
\end{figure}

The BOgen system comprises three main components: 1) Generative AI for 3D designs, 2) an exploration map based on the reduced latent space of automatically generated 3D designs, 3) a PBO model that infers user preferences in the reduced latent space, and 4) an interface that enables designers to effectively generate and explore part-level designs. Briefly, BOgen maps user-synthesized chairs to those not selected for a reduced latent space (\textit{z1},\textit{z2} in Figure \ref{fig:2}). Based on this information, the PBO infers user preferences. Subsequently, the points
sampled by the PBO are decoded and passed through the 3D Generative AI model (in this study, SALAD),
resulting in chair design generation.

\subsection{Dimensionality Reduction of SALAD Extrinsic Latent using VAE}
\begin{figure}[ht]
  \centering
  \includegraphics[width=\linewidth]{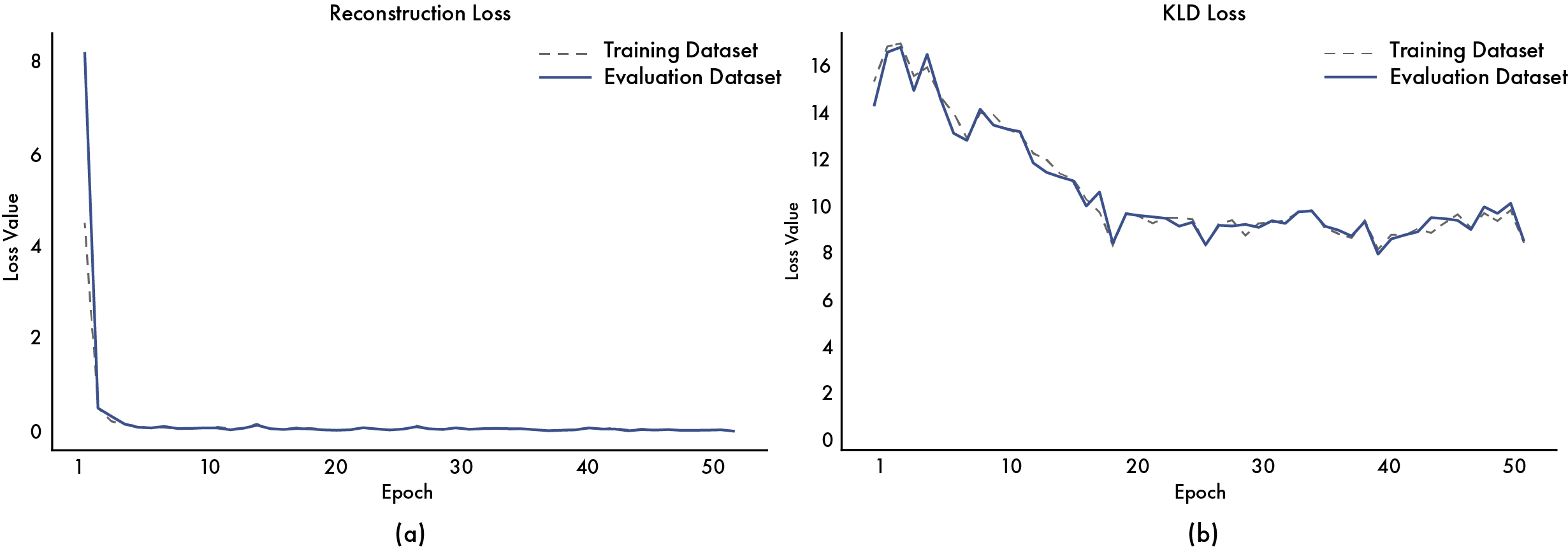}
  \caption{VAE training and evaluation loss: (a) reconstruction loss; and (b) KLD loss.}
  \label{fig:3}
\end{figure}

The extrinsic latent component of the SALAD consists of 16 parts, each with 16 dimensions. Although these latents effectively represent the broad shape of intricate 3D objects within a condensed dimension, adjusting each of the 16 parts to match a designer’s specifications is both laborious and time-intensive. Furthermore, visualizing this 256-dimensional design space for designer exploration poses several challenges. Therefore, we utilized a VAE
to map the 16x16-dimensional latent to 2D without losing detailed information and to generate an extrinsic latent based on the 2D points. We mapped the 16×16-dimensional extrinsic latents to an explorable dimension (n=2) to create an exploration map (\textit{z1},\textit{z2} in Figure \ref{fig:2}), referring to points on this map as ‘2D latent points’. For the training and evaluation datasets, 58,750 extrinsic latent variants from the SALAD prompt training dataset were used, with
52,992 for training and 5,758 for evaluation ]\cite{koo2023salad}. In the VAE encoder, the Kullback-Leibler divergence (KLD) is used as the loss function to measure the difference between the target and trained distributions \cite{hyun2018balancing}. The decoder’s loss function employed the mean squared error (MSE) to represent the similarity between the decoded data and the ground truth. The training session took 50 epochs (Figure \ref{fig:3}). Consequently, the VAE reconstruction loss on
the evaluation dataset was approximately 0.015, indicating that the generated data closely resembled the ground truth (i.e., extrinsic latent data). The data distribution mapped to the reduced 2D design space showed dense areas for chair types, such as armchairs and dining chairs (Figure \ref{fig:4}-a, b). Moving to the top left, many decorative chairs, folding chairs, and stools were distributed (Figure \ref{fig:4}-c, d), followed by unconventional chairs (Figure \ref{fig:4}-e). We obtained 1,981 characteristic data points from a dataset of 58,750 training and evaluation datasets using the kmeans++ algorithm, which was then utilized as the final dataset.

\subsection{Preferential Bayesian Optimization}
\subsubsection{PBO Algorithm}
BOgen employs a PBO to predict and provide the user with the desired data based on their preferences.
However, estimating the designer’s desired part from existing extrinsic (16 × 16 dimensions) or intrinsic latent (16
× 512 dimensions) is computationally inefficient. Therefore, to solve this problem, we propose a method that
applies a PBO based on a reduced exploration map via a VAE to estimate and suggest a user’s desired design in
real-time. The PBO used in our study followed that of Koyama and Goto \cite{koyama2022bo}. First, the objective function of the
PBO is the function g which predicts the user’s desired design. For instance, if the user is presented with n chairs
and prefers chair number 4, we represent this observation as \(\mathbf{d} = [x_4 > x_1, x_2, x_3, x_5, \ldots, x_n]\). This means \(x_4\) represents the 'preferred design', while the rest are 'other designs.' With multiple observations of \(\mathbf{d}\), it is denoted as \(\mathbf{D} = [\mathbf{d}_1, \mathbf{d}_2, \ldots, \mathbf{d}_n]\). For each data point \(x_i\), a preference or 'goodness' value, denoted by \(g_i\), is assigned, where \(i\) ranges from 1 to \(N\). These values, \(g_i\), are represented as \(\mathbf{g} = [g_1, \ldots, g_N]^\top\). For the preference-based objective function \(g\), preferential modeling uses the data created or preferred by the user to model \(\mathbf{d}\), the relative preference among the remaining provided data, using the Bradley-Terry-Luce model \cite{bradley1952rank}:
\begin{equation}\label{eq1}
P(d|g) = \frac{exp(g^{(i)})}{\sum_{j=1}^N exp(g^{(j)})}.
\end{equation}
The Bradley-Terry-Luce (BTL) model was used to model the relative preferences among the given data based on user preferences. The likelihood of multiple observation data, given the preference values \( \mathbf{g} \), is calculated as the product of individual likelihoods, represented by \( P(\mathbf{D} | \mathbf{g}) = \prod_i P(\mathbf{d}_i | \mathbf{g}) \). However, the \( g(x) \) values for unobserved data points \( x \) remain unknown. To predict these values, a Gaussian process (GP) was used, which assumes that all data points share a common probability space \cite{seeger2004gaussian}. Based on this assumption, the relative preference \( \mathbf{d} \) calculated through the BTL is used to estimate \( g(x) \), reflecting the user’s desired design. The maximum a posteriori (MAP) methodology is then applied to obtain an optimal estimate for these \( g(x) \) values.
\begin{equation}\label{eq2}
g^{MAP} = argmax P(g|D) = argmax P(D|g)P(g).
\end{equation}
Essentially, GP estimates various possible probability distributions for \( g(x) \) based on the observed data, and MAP identifies the distribution with the highest posterior probability that most closely aligns with the observed data. Specifically, in Eq. 2, \( P(g|\mathbf{D}) \) represents the posterior probability of \( g(x) \) values, given the observed data \( \mathbf{D} \), expressed as the product of \( P(\mathbf{D}|g) \) and \( P(g) \). \( P(\mathbf{D}|g) \) denotes the conditional probability of \( \mathbf{D} \) given the \( g(x) \) values (i.e., the probability of occurrence of \( \mathbf{D} \) when \( g(x) \) values are given), and \( P(g) \) represents the prior distribution of \( g(x) \) values, which, under GP assumptions, is a Gaussian distribution \cite{koyama2022bo}. Using the estimated \( g(x) \) values, the predicted distribution of an unseen data point \( x \) can be calculated as follows:
\begin{equation}\label{eq3}
g(x) \sim N (\mu(x), \sigma^{2}(x)).
\end{equation}
This stems from the Gaussian Process (GP) model, where \( g(x) \) is defined by a normal distribution. The mean, \( \mu(x) \), represents the predicted user preference for design option \( x \), and the variance, \( \sigma^2(x) \), quantifies the uncertainty associated with this prediction \cite{koyama2022bo}. Specifically, a high \( \mu(x) \) in the normal distribution of a particular data point \( x \) implies a high \( g(x) \) value, indicating a high probability of user preference. Conversely, a large \( \sigma^2(x) \) value suggests greater uncertainty in GP’s prediction, indicating less confidence about that particular data point \( x \). The Acquisition function used to provide suggestions to the user based on these estimated \( \mu(x) \) and \( \sigma(x) \) employs the Gaussian process upper confidence bound (GP-UCB) \cite{srinivas2012information}:
\begin{equation}\label{eq4}
a^{GP-UCB} = \mu(x) + \beta \sigma(x).
\end{equation}

The GP-UCB is an acquisition function based on a Gaussian process. In our study, the \(\beta\) parameter in GP-UCB (Eq. 4) is a hyperparameter that determines the level of exploration and exploitation in the PBO model, and we used a value of 0.5 to ensure balanced exploration and exploitation \cite{koyama2022bo}. It recommends optimal suggestion \( x \) to the user using the mean and variance of the preference-based objective function. In other words, PBO suggests the \( x \) points with the highest acquisition values. To generate multiple \( x \) samples, the batch BO method, which performs Eq.4 \( k \) times (\( k=16 \)), was used to create multiple samples \cite{schonlau1998global}.

\subsubsection{Preferential Bayesian Optimization for BOgen}
\begin{figure}[ht]
  \centering
  \includegraphics[width=\linewidth]{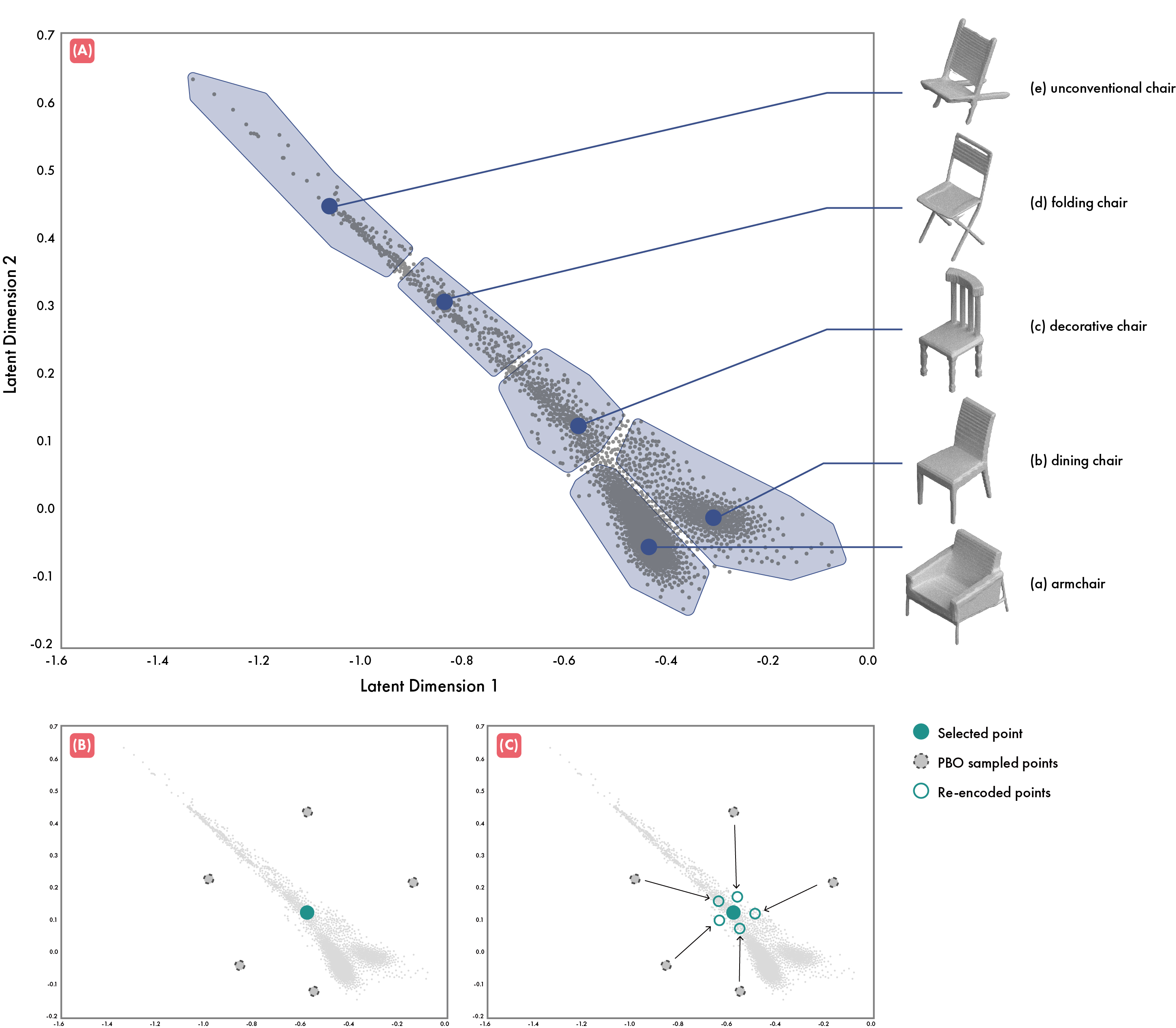}
  \caption{Data distribution of exploration map and example of the process of mapping PBO generated samples: (A)
five clusters divided by chair types; (B) PBO sample points and selected point; and (C) re-encoded points with
skip information of selected point.}
  \label{fig:4}
\end{figure}

\begin{figure}[ht]
  \centering
  \includegraphics[width=\linewidth]{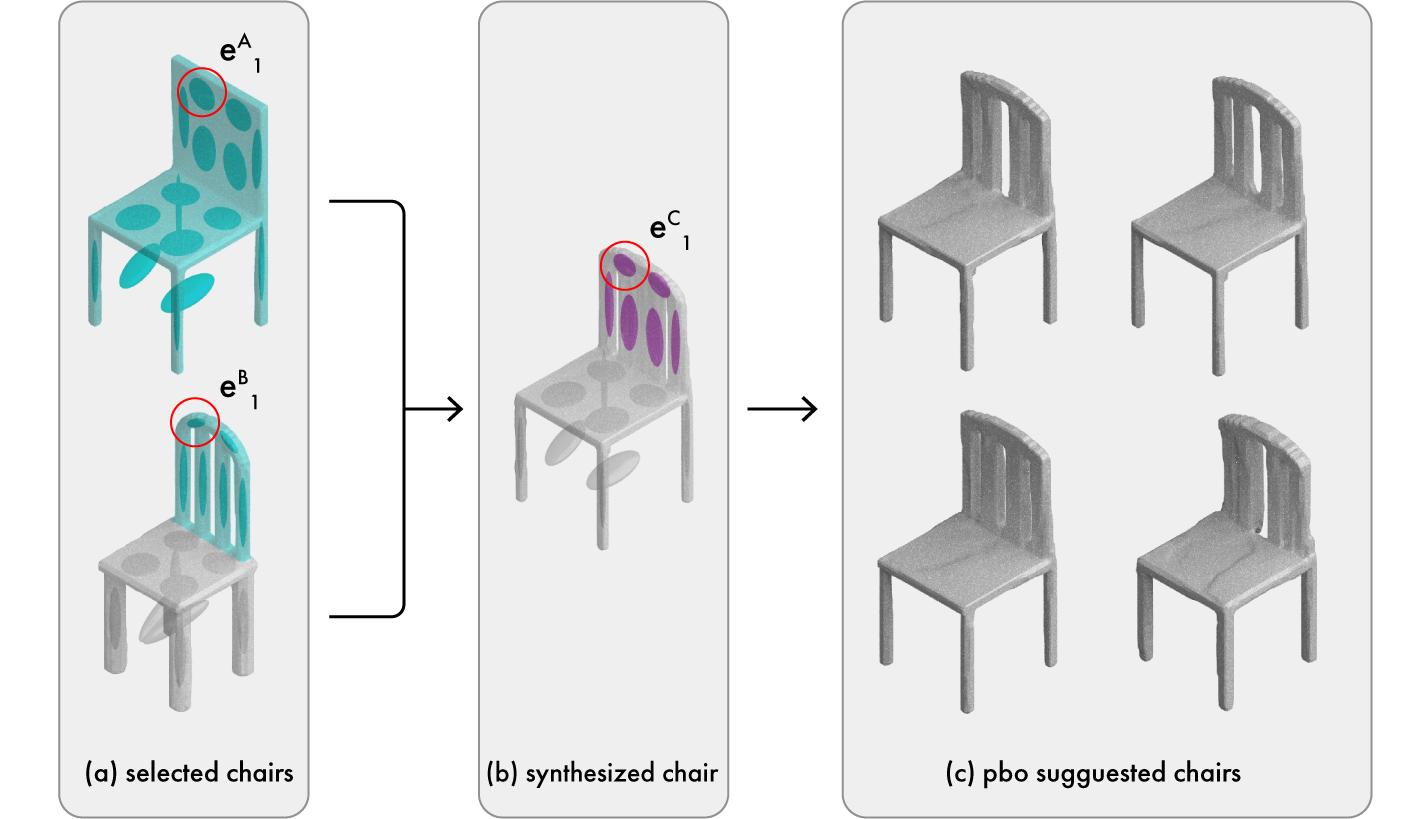}
  \caption{Example of PBO generated chairs based on same skip information and different sampled points: (a)
selected chairs; (b) synthesized chair; and (c) PBO suggested chairs.}
  \label{fig:5}
\end{figure}

\textbf{Generating samples with VAE.} We set the sampling boundary of the PBO to encompass the 2D latent-space range of the VAE. The range for the first latent dimension was set between 0 and 1.5, while the second dimension’s range was between -0.1 and 0.7. The 2D latent points sampled by the PBO were decoded into extrinsic latent points by utilizing skip information from the user-selected chair. Once decoded, these extrinsic latent points were reencoded, and the resultant 2D latent points were appended to the exploration map. In essence, the 2D latent points originally sampled by the PBO underwent a decoding process with the chosen skip information, resulting in the generation of an extrinsic vector. This vector is then re-encoded and integrated into an exploration map. As illustrated in Figure \ref{fig:4}-B, and C, when the PBO samples points based on user preference data (Figure \ref{fig:4}-B), these
points are decoded through the VAE. The skip information from the selected points was re-encoded and plotted around the originally chosen points (Figure \ref{fig:4}-C). Consequently, the PBO creates chairs with unique details by drawing from parts recently generated by the user (Figure \ref{fig:5}).

\textbf{Updating Preference-based Objective Function.} We regard the 2D latent points of chairs for which users have explicitly shown preferences or requested PBO suggestions (i.e., want to see similar designs) as the preferred design. The 2D latent points of all the other chairs were considered as the other designs. If a chair preferred by the user or for which a PBO suggestion was requested was a combination of more than one chair, those chairs were excluded from the other designs. In addition, the original 2D latent points sampled by the PBO (i.e., points before being encoded) were updated as other designs and excluded from the next update. This was performed to prevent the PBO from continuously sampling the same points.

\subsection{BOgen User Interface}
\subsubsection{User Interface Main Screen}
\begin{figure}[ht]
  \centering
  \includegraphics[width=\linewidth]{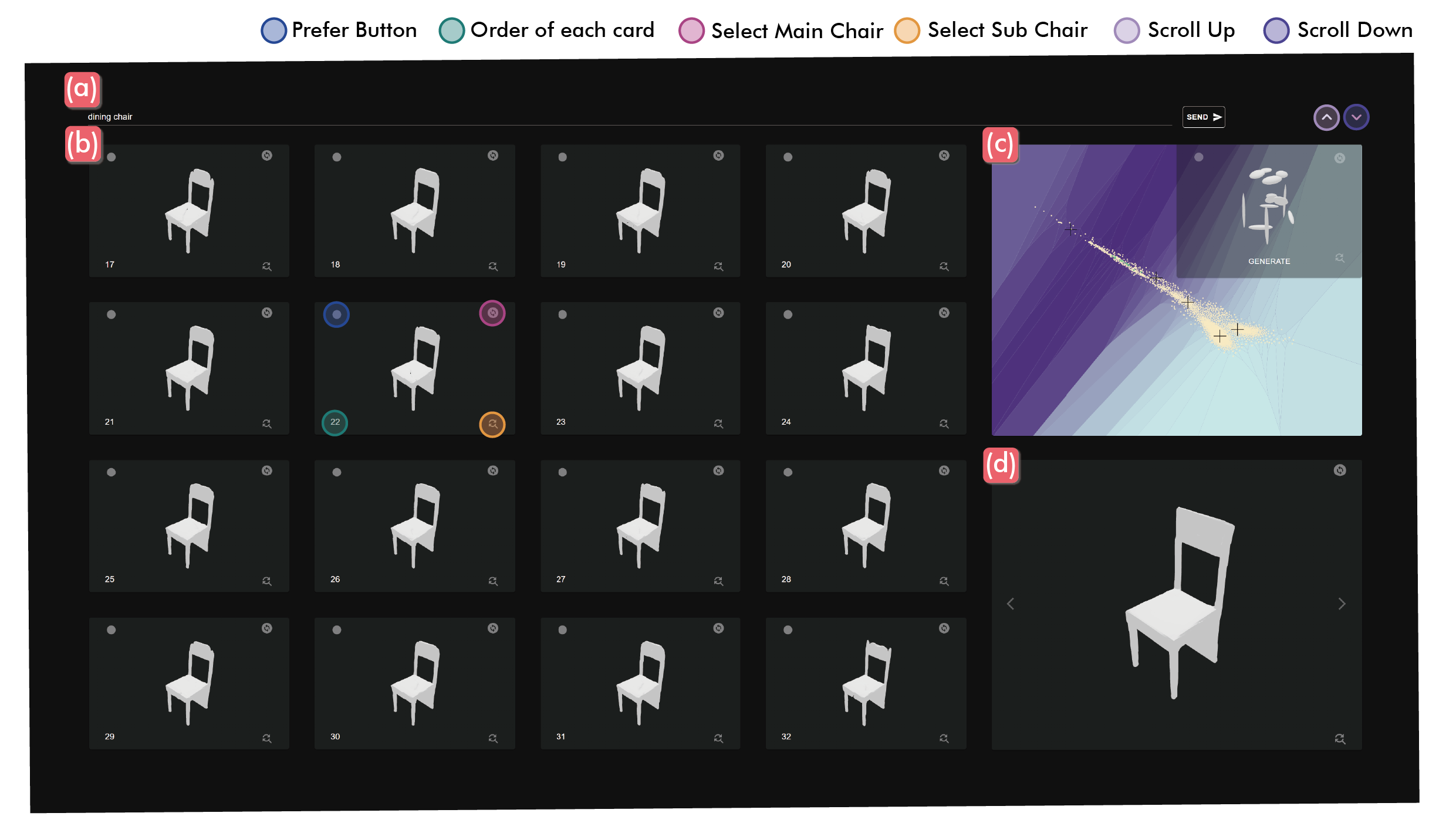}
  \caption{BOgen System Overview: (a) prompt input field; (b) generated chairs by prompt or PBO; (c) exploration
Map; and (d) added synthesized chair}
  \label{fig:6}
\end{figure}

The BOgen system interface is shown in (Figure \ref{fig:6}). Initially, users could input their desired chair into the prompt (Figure \ref{fig:6}-a), and the results could be viewed 16 times simultaneously. The sequence of each card is displayed at the bottom left. Additionally, a 2D exploration map located at the top right facilitates the navigation of designs, allowing users to explore beyond those shown on the cards (Figure \ref{fig:6}-c). Users can mark the designs of interest using a button at the bottom left of each card (Figure 6-b). Such marked designs are considered the preferred designs and are excluded from subsequent updates and other designs. Upon selecting the main and subdesigns through the generated chairs and map, a screen pops up, allowing for the synthesis of different parts. The resulting design can then be saved (Figure 6-d). Scrolling down reveals the next 16 cards, displaying chairs sampled by the
PBO based on the user's recent synthesis. The VAE latent of this design is deemed the preferred design, whereas the VAE latent of other designs that are not part of the synthesis is categorized as other designs. 

\subsubsection{Synthesis Screen}
\begin{figure}[ht]
  \centering
  \includegraphics[width=\linewidth]{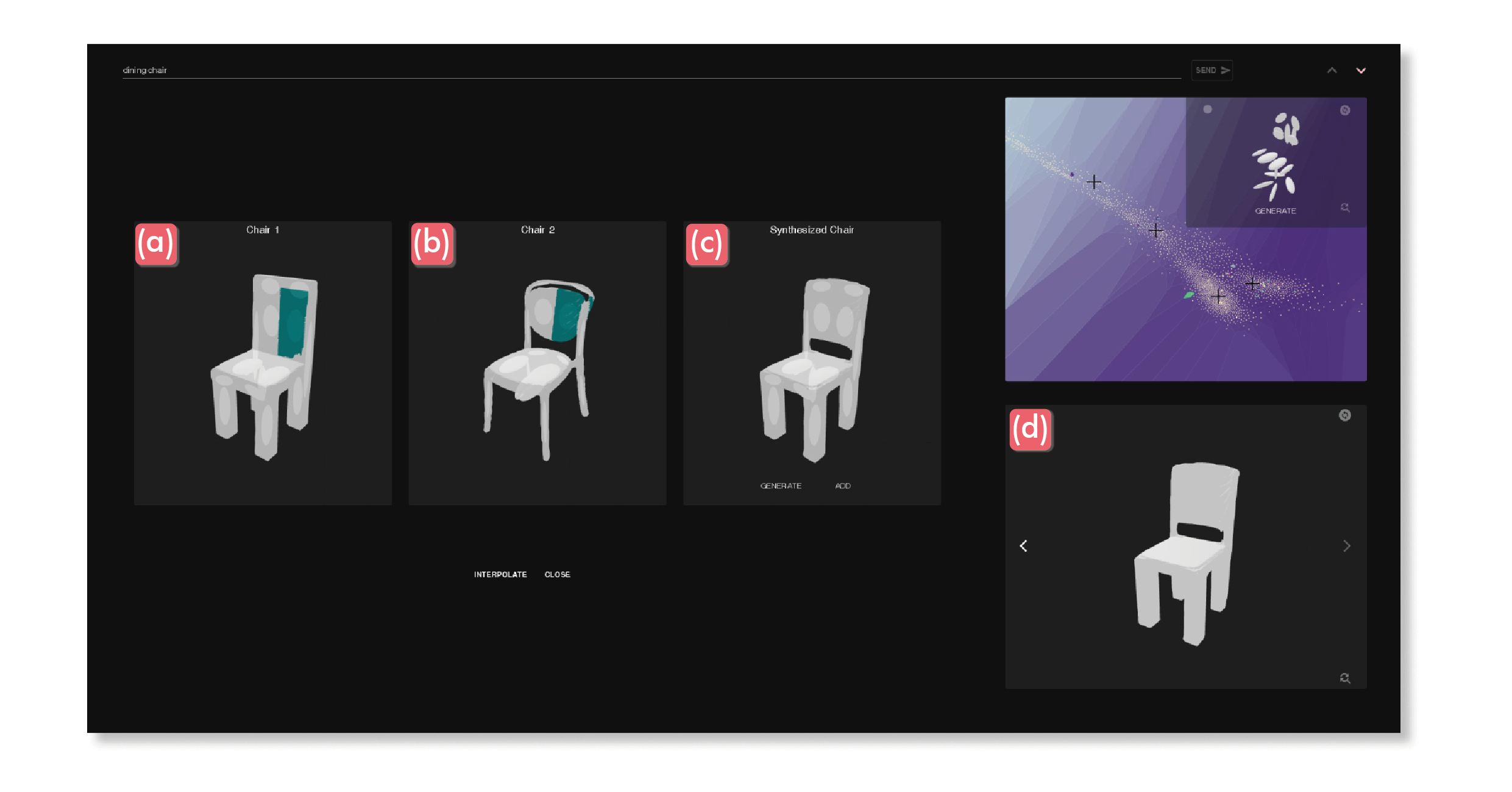}
  \caption{BOgen chair synthesizing interface.}
  \label{fig:7}
\end{figure}

When the main chair and subchair are selected on the BOgen main screen, a synthesis screen appears (Figure \ref{fig:7}). Within the synthesized chair, parts of the main chair (Figure \ref{fig:7}-a) and subchair (Figure \ref{fig:7}-b) can be selected individually. Upon pressing the interpolate button, changes in the 3D Gaussian function based on the main chair were observed. For example, if a user interpolates the 1st part of the 3D Gaussian of the main chair A (\(e^{A} _{1}\)) with the 1st part of the 3D Gaussian of sub-chair B (\(e^{B} _{1}\)), the newly synthesized 1st part of the 3D Gaussian of chair C (\(e^{C} _{1}\)) is \((e^{A} _{1} + e^{B} _{1}) / 2\). The 3D Gaussians of the remaining parts remain unchanged and are the original ones from the main chair A. By pressing the generate button, the shape of the synthesized chair can be previewed (Figure \ref{fig:7}-c), and by clicking on "Add,” it can be saved to the card located at the bottom right of the main screen (Figure \ref{fig:7}-d).

\subsubsection{Exploration Map}
\begin{figure}[ht]
  \centering
  \includegraphics[width=\linewidth]{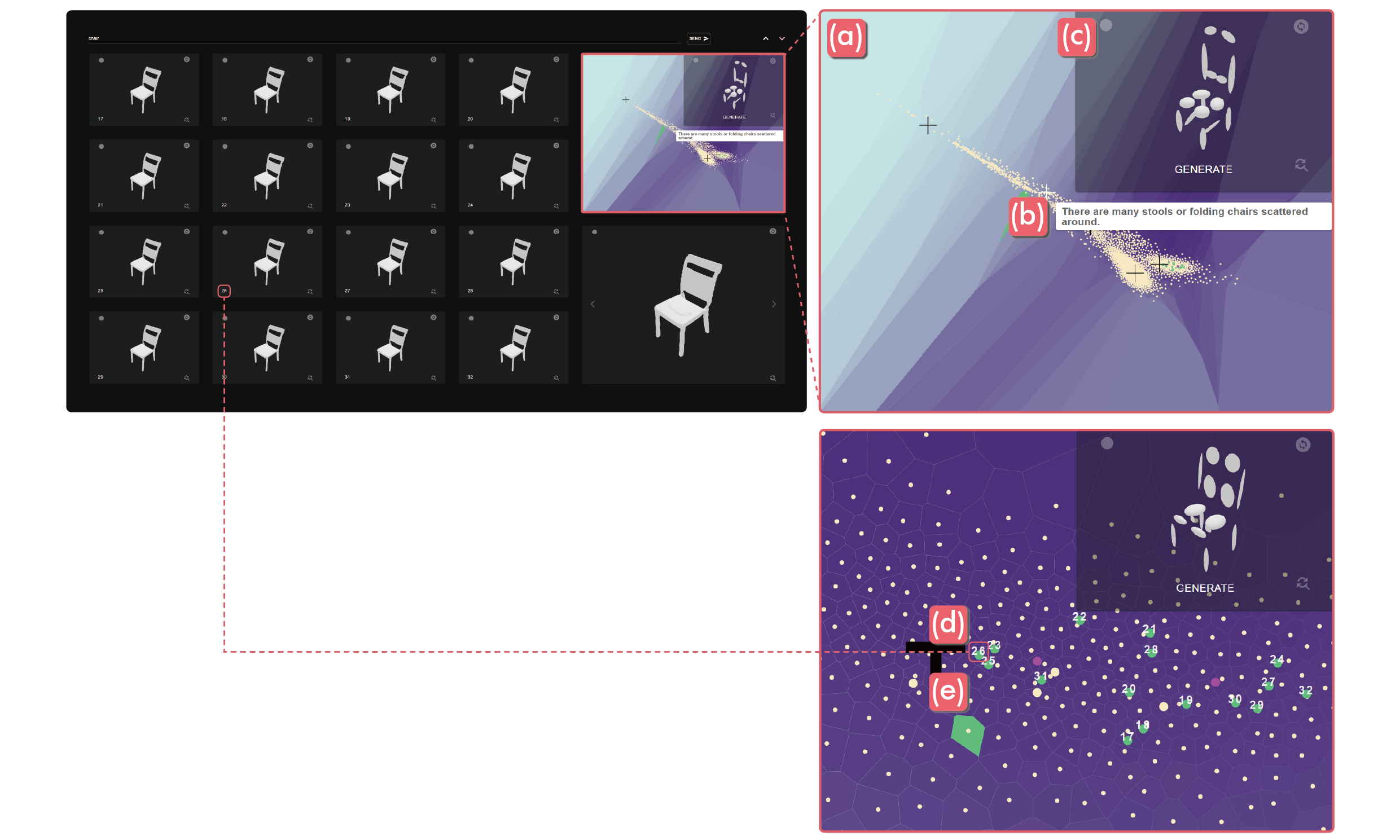}
  \caption{Exploration Map: (a) exploration map; (b) brief explanation; (c) 3D gaussian of a point ;(d) order of each card; and (e) exploration area mark.
}
  \label{fig:8}
\end{figure}

Designs can be explored through the “exploration map” located at the top right of the main screen. The color of the area was updated (from purple to sky blue) based on the probability of the chair being selected (Figure \ref{fig:8}-a). Each area of the map is marked with a "+" sign to indicate the characteristics of the designs distributed (Figure \ref{fig:8}-b). Hovering over a point representing a design on the map displays it in a 3D Gaussian form at the top right (Figure \ref{fig:8}-c). Designs explored on the map can also be selected as either the main or sub-design and synthesized in the same manner as before. The 16 chairs displayed on the cards were numbered (marked) on the map (Figure \ref{fig:8}-d). In addition, clicking on the area of the map highlights it as fluorescent lime, indicating the exploration area (Figure \ref{fig:8}-e).

\section{IMPLEMENTATION AND RESULTS}
\subsection{Experimental Design}
\begin{figure}[ht]
  \centering
  \includegraphics[width=\linewidth]{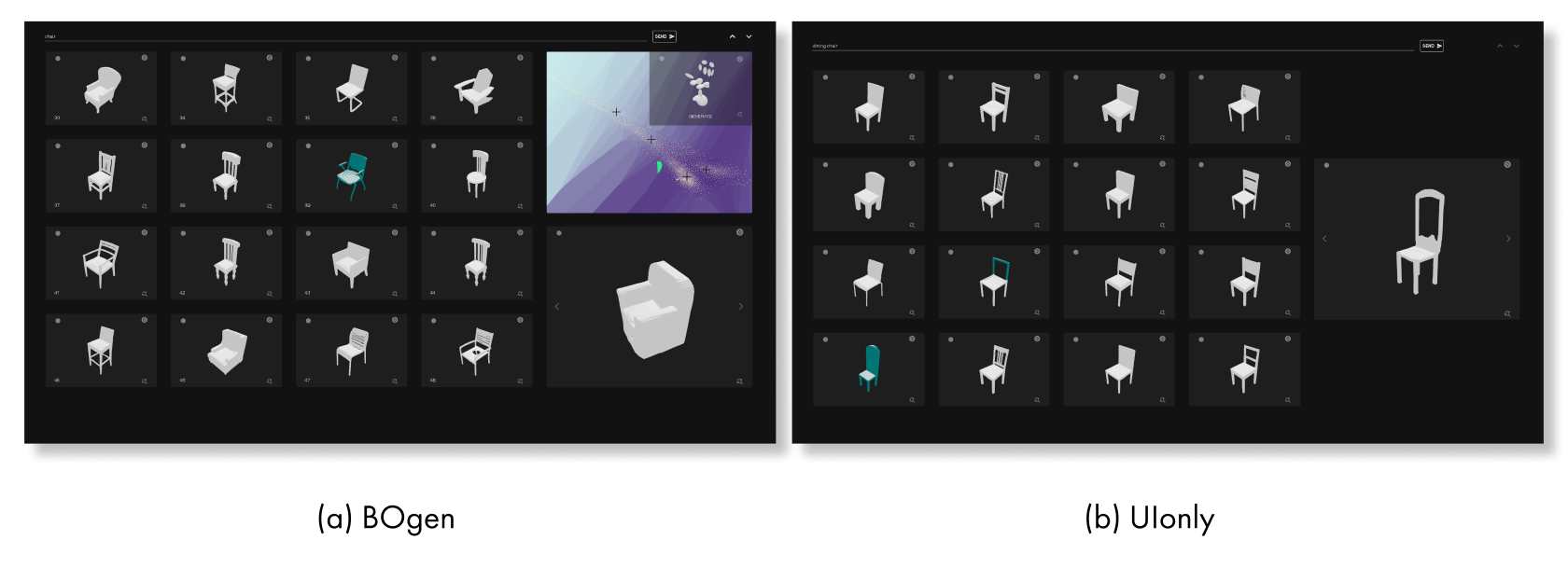}
  \caption{System Interface for (a) BOgen; and (b) UIonly}
  \label{fig:9}
\end{figure}

To evaluate BOgen’s performance, we recruited 30 experimental participants (including 20 females), all of whom had majors related to design or had professional design experience in the field. (Age: Mean = 23.03; Max = 27; and Min = 19). The experiment followed a within-subject design and proceeded in the following order: introduction, consent, and tutorial (about 10 minutes) → Brief 1 (about 30 minutes) → Brief 2 (about 30 minutes) → In-depth interview (about 20 minutes). The participants performed two tasks (BOgen and UI only). Here, the UIonly served as a baseline interface for comparison, excluding the exploration map and PBO-based recommendation features from the BOgen system (Figure \ref{fig:9}). The remaining synthesis function and prompt-based
creation used the same model. Each exploration task required participants to use both systems to find or design chairs suitable for specific spaces, such as living rooms or offices, as depicted in the images provided. They were given a maximum of 30 minutes for each task, with the option to conclude earlier if they generated or identified a design outcome that met their satisfaction. The tasks were performed in an even order based on the system, and
the experiment was conducted using design briefs with similar details. The system used in the experiment was developed using Python 3.9. All participants used a high-performance client-server system (client: React on Windows OS with Intel Core i9 10980XE, 64 GB RAM; server: Python Flask on Linux OS with AMD Ryzen Threadripper Pro 3995WX, 256 GB RAM). During each experiment, the extrinsic and 2D latent points of all the generated and explored chairs were logged. After each session, a 7-point scale survey was conducted to assess the influence of the system on the design process. After all the sessions were completed, we collected opinions on the overall experience of both systems through interviews.

\subsection{Experimental Analysis Metrics}
In our experimental analysis of BOgen, we focused on four key metrics along with surveys and in-depth interviews for a comprehensive evaluation. Four quantitative analysis metrics were utilized: 1) the mean uncertainty evaluates the model’s ability to accurately capture and adapt to user preferences, indicating the model's confidence level in its predictions; 2) the mean probability of the preferred chair assesses how well the system identifies designs that align with the user's preferences, reflecting the system’s predictive accuracy; 3) the explored area captures the extent of the user's exploration within the design space, highlighting the breadth of their design exploration; and 4) the number of clusters indicates the density and distribution of user design exploration, providing insights into the patterns of their exploration behavior. We performed paired-sample t-tests on the metrics that underwent statistical analysis.

\subsubsection{Mean Uncertainty}
We evaluated all the 2D latent points generated by the user, encompassing the initial 1,981 points, as the information space. We established the average value of uncertainty in that space, denoted as \( \sigma^2(x) \), as the mean uncertainty. The mean uncertainty serves as a crucial metric, reflecting the model's confidence in capturing and adapting to the user's evolving preferences over time. Calculating the mean uncertainty involved deriving the average uncertainty across all information spaces after executing a PBO update. Therefore, a lower mean uncertainty indicates a high level of confidence in the model's predictions within the information space. Specifically, for the i-th PBO update, the uncertainty for the information space can be expressed as the average of \( \sigma^2(x_i) \) over all \( x \) in the space. This is computed as mean uncertainty \( = \frac{1}{N} \sum \sigma^2(x_i) \), where \( N \) represents the total number of points in the information space.

\subsubsection{Mean Probability of Preferred Chair}
The mean probability of the preferred chair represents the average of the prior probabilities (i.e., \( \mu(x) \)) of the preferred designs, representing how well the system aligns with the user's preferences. A high mean probability value reflects a strong alignment with user preferences, indicating the system's effectiveness in identifying preferred designs. Suppose only one design is preferred while the user uses the system, and it results in a prior probability value of 0. In that case, it suggests that the system may not fully understand the user’s preferences yet. Mean probability is calculated as the average of the \( \mu(x) \) values for each preferred design prior to its PBO update. For example, if the designer preferred \( x_1 \) design in 4th PBO iteration state, the \( \mu(x_{4_1}) \) is the prior probabilities. This can be mathematically expressed as mean probability \( = \frac{1}{N} \sum \mu(x_i) \), where \( N \) represents the total number of preferred designs.

\subsubsection{Explored Area}
We defined the 2D latent points of the chairs that users searched for, generated, and hovered over as the 2D latent points explored by the users. The explored area is a significant indicator of a user's exploratory behavior, showing the breadth and diversity of the design elements investigated. The explored area is determined by the convex hull of the 2D latent points, representing the breadth of the design explored by the user on the exploration map. Mathematically, if we denote the set of explored 2D latent points as \( X = \{x^1, x^2, \ldots, x^n\} \), where each \( x^i \) is a point in the latent space, the area of the explored region, \( A \), can be calculated as the area of the convex hull formed by the points in \( X \). This is expressed as \( A = \text{Area}(\text{convex hull} (X)) \).

\subsubsection{Number of Cluster}
To measure the density of the 2D latent points explored by the users, we utilized the DBSCAN algorithm. DBSCAN is renowned for clustering based on point density. The number of clusters identified by DBSCAN is critical because it provides insight into the patterns and structures within the user's exploration, indicating areas of concentrated interest. Before using DBSCAN, one must determine the \( \varepsilon \) and MinPts value. The \( \varepsilon \) value serves as a parameter to find neighboring points within the \( \varepsilon \) distance from a given data point. Too small a value categorizes many data points as noise, while too large a value can merge different clusters. A common approach to finding an appropriate \( \varepsilon \) is to identify the "elbow point" in a k-neighbor graph. This point corresponds to a sudden change in the graph’s slope, where the curvature is maximized. By setting the \( \varepsilon \) based on this "elbow point," one can optimize clustering performance. Based on the 2D latent points explored in each system by users, we set an elbow point with a maximum curvature value of 0.05 or less. Ultimately, the average of these 60 elbow points (number of experiment participants * number of systems) was set to 0.039 for DBSCAN’s \( \varepsilon \). MinPts was set to 3.

\subsubsection{Survey and In-depth Interview}
To assess if BOgen delivered designs with the user's preferred parts, assisted in design development and navigation,
and provided useful designs, we surveyed the following four questions on a 7-point Likert scale: Q1) The system
effectively suggested designs containing my desired parts; Q2) The system effectively clarified my design goals;
Q3) The system assisted in navigating towards my preferred design direction; and Q4) The system provided useful
design suggestions. Additionally, an in-depth interview inquired about the impact of each system on the design
process, the pros and cons experienced during system usage, and their opinions.

\subsection{Results and Discussions}
Based on the experimental results, when compared with the UIonly system, BOgen exhibited three main findings:
1) guiding design space navigation for design idea refinement, 2) recommendations for designers to find prominent
design parts, and 3) facilitating designers to explore dense and diverse design spaces.

\subsubsection{Guiding Design Space Navigation for Design Idea Refinement}
\begin{figure}[ht]
  \centering
  \includegraphics[width=\linewidth]{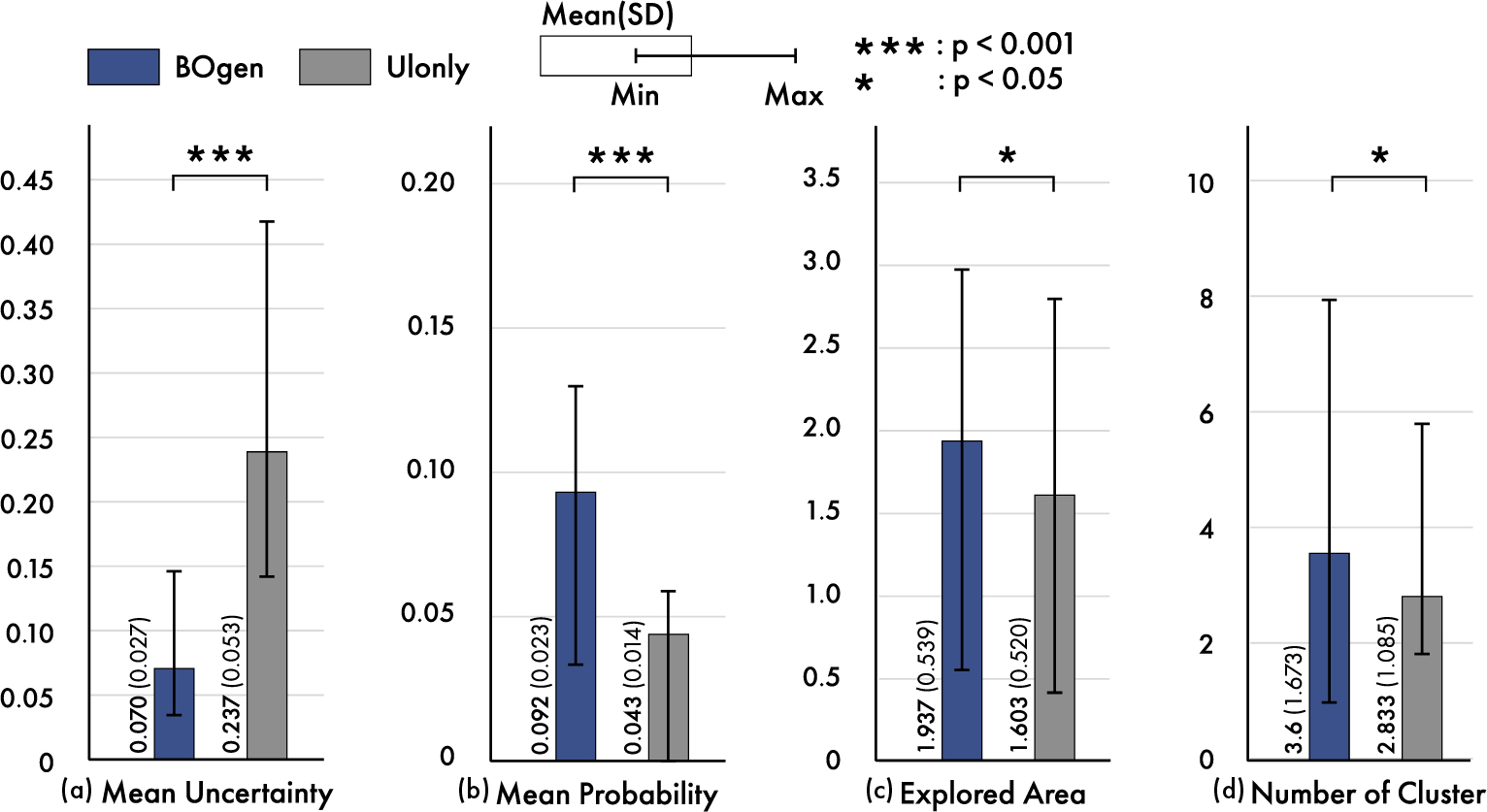}
  \caption{Results of (a) mean uncertainty; (b) mean probability; (c) explored area; and (d) number of clusters.}
  \label{fig:10}
\end{figure}

\begin{figure}[ht]
  \centering
  \includegraphics[width=\linewidth]{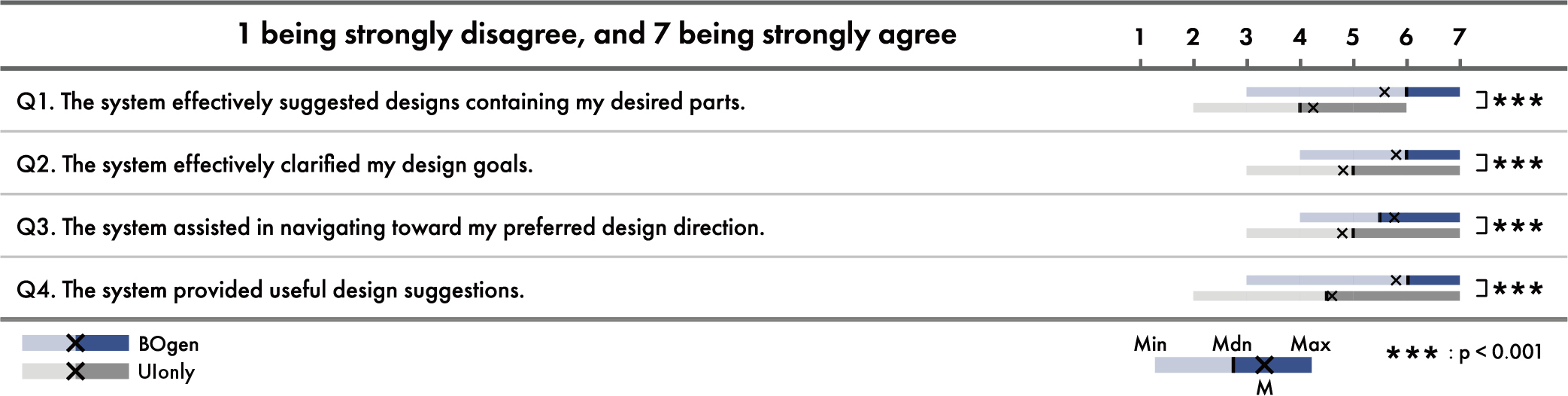}
  \caption{Results of survey.}
  \label{fig:11}
\end{figure}

During the design process, designers continuously refine their design ideas and navigate the design space to search for and integrate the desired design parts. Therefore, effectively guiding the design space to infer the designer’s preference, especially those with desired parts, is a crucial task. To accomplish this task effectively, a system needs to better capture user preferences and discern the preferred designs. We employed the mean uncertainty and mean probability metrics to measure the system’s capability to reflect user preferences and identify preferred designs. Consequently, BOgen displayed a significantly lower mean uncertainty (BOgen: 0.070; UIonly: 0.237; \( p<0.001 \); Figure \ref{fig:10}-a) and a significantly higher mean probability (BOgen: 0.092; UIonly: 0.043; \( p<0.001 \); Figure \ref{fig:10}-b) compared to the UIonly system. This indicates that BOgen captured user preferences more accurately and had greater confidence in determining the chairs preferred by the users. Furthermore, for survey questions Q2 and Q3, BOgen scored significantly higher (Figure \ref{fig:11}; Q2: MeanBOgen = 5.8, MeanUIonly = 4.833, \( p<0.001 \); Q3: MeanBOgen = 5.766, MeanUIonly = 4.8, \( p<0.001 \)). The majority of users (90\%) expressed that they could better navigate towards their desired design direction and refine their ideas with the aid of BOgen’s exploration map and PBO-based recommendations. Some notable examples of user feedback include the following:
\begin{itemize}
    \item P1: \textit{"With the system (BOgen), the map offered unexpected design inspirations and shapes. I could view something unique on the map and then scroll to see the various variations with (PBO) suggestions, which was helpful."}
    \item P7: \textit{"There were usable designs in the (PBO) recommendations. The map, with its color information, helped me look at both areas of interest and those that I had not previously considered, aiding in synthesizing elements and invoking images."}
    \item P24: \textit{"I had a design intent, and the map distribution was helpful when specifying it. I was looking for a unique shape, and it was great not having to search but to have a guide. A similar recommendation function (PBO) is also helpful for seeking details. The ideation process benefited from observing several similar details."}
\end{itemize}
In summary, these results demonstrate that BOgen captured user preferences more effectively, made accurate predictions about user-preferred designs, and guided design-space navigation for refining design ideas.

\subsubsection{Recommending to Designers to Find Prominent Design Parts}
\begin{figure}[ht]
  \centering
  \includegraphics[width=\linewidth]{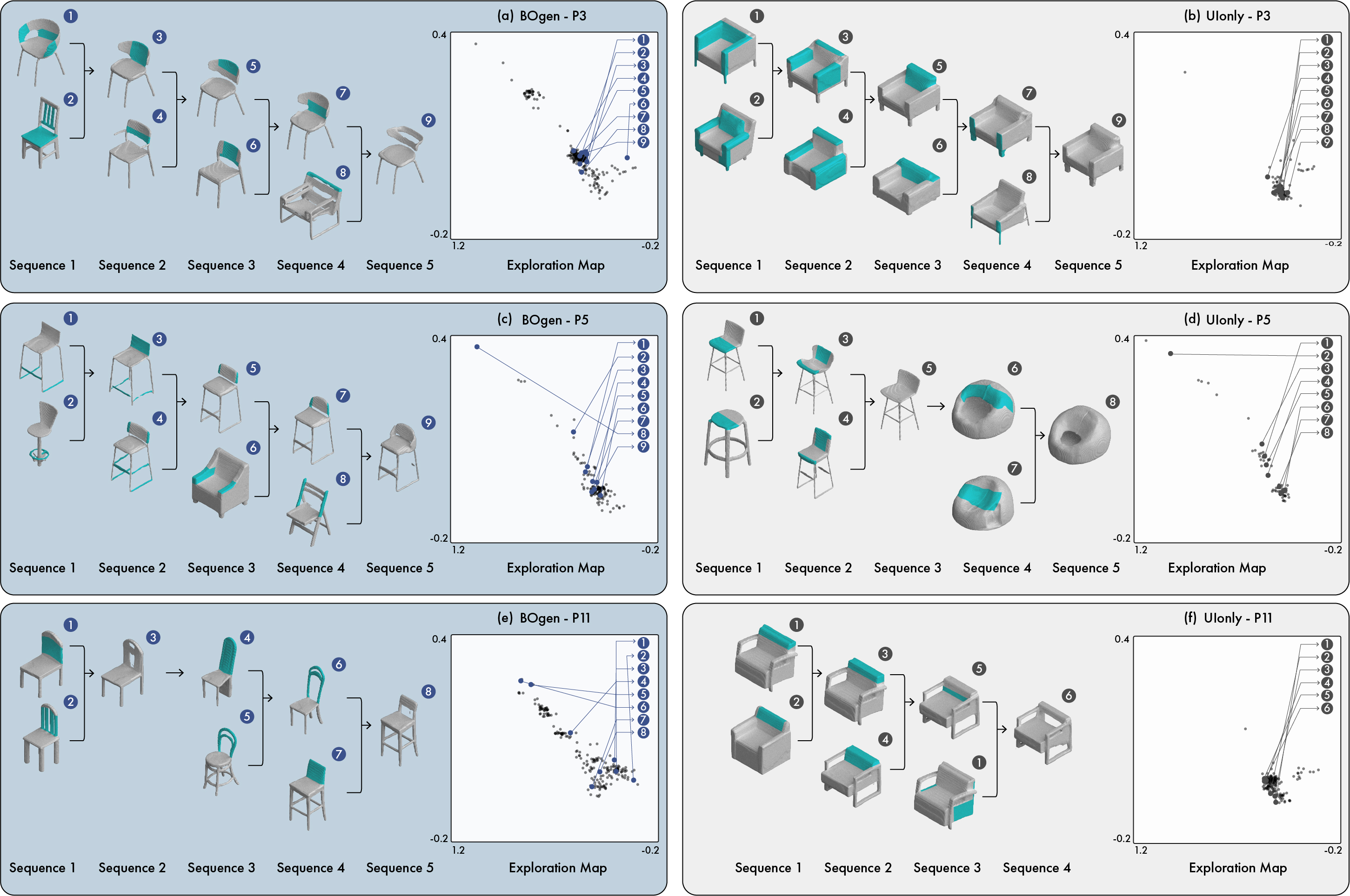}
  \caption{Sequential exploration path of BOgen and UIonly: (a) BOgen of P3; (b) UIonly of P3; (c) BOgen of P5;
(d) UIonly of P5; (e) BOgen of P11; and (f) UIonly of P11.}
  \label{fig:12}
\end{figure}

Designers’ intentions evolve as they progress through the design process. Hence, it is essential to develop designs that respond to these changing needs. In this regard, BOgen scored statistically significantly higher in surveys Q1 and Q4 compared to the UIonly system (Figure \ref{fig:11}; Q1: \textit{MeanBOgen} = 5.6, \textit{MeanUIonly} = 4.233 , p<0.001 ; Q4: \textit{MeanBOgen} = 5.8 , \textit{MeanUIonly}= 4.6 , p<0.001). This indicates that BOgen more effectively recommended designs that included the desired design parts and provided valuable suggestions to users. To further investigate these results, we analyzed users’ sequential exploration paths. For instance, P3 was searching for a chair with a rounded backrest. After searching "round table chair", P3 explored the 2D latent points of the resulting chairs. P3 found a chair corresponding to sequence 1 in figure \ref{fig:12}-a and combined it with the original search result, creating a chair in sequence 2. After that, they continued exploring and merging chairs. Subsequently, P3 searched "backrest chair" and combined the resulting design with their previously developed design. During the interview, P3 mentioned, "(In BOgen) I was looking for a chair with a rounded backrest. I explored around the dots on the map ... I went on to find the desired part by looking at the dots on the map." On the other hand, with the UIonly system, P3 said, "I had a hard time searching because I could only think of a few keywords. When I searched for a sofa, it didn’t match my desired design, so I had to synthesize more." Observing Figure \ref{fig:12}-c, P5 started by searching for a stool, and while exploring within those results, in sequence 3, P5 merged their design with a chunky armchair and then chose a chair based on the PBO recommendation. In sequence 4, P5 found a folding chair on the exploration map and combined it, receiving another PBO recommendation and selecting a chair in sequence 5. P5 shared in an interview, "Because I only knew a limited type of chairs, I found using the map in the system (BOgen) more helpful than just keywords (UIonly)." Lastly, as seen in Figure \ref{fig:12}-e, P11 began by exploring the 2D latent points of their initial search result on the map. Particularly in sequence 3, P11 found desired chairs on the exploration map and kept developing their design. P11 said, "There were many unexpected discoveries from the map or (PBO) recommendations. The process was faster, and I could pull out many designs." To sum up, BOgen was superior to UIonly in effectively suggesting the design elements users were looking for and presenting them with prominent designs.

\subsubsection{Facilitating Designers to Explore Dense and Diverse Design Space}
\begin{figure}[ht]
  \centering
  \includegraphics[width=\linewidth]{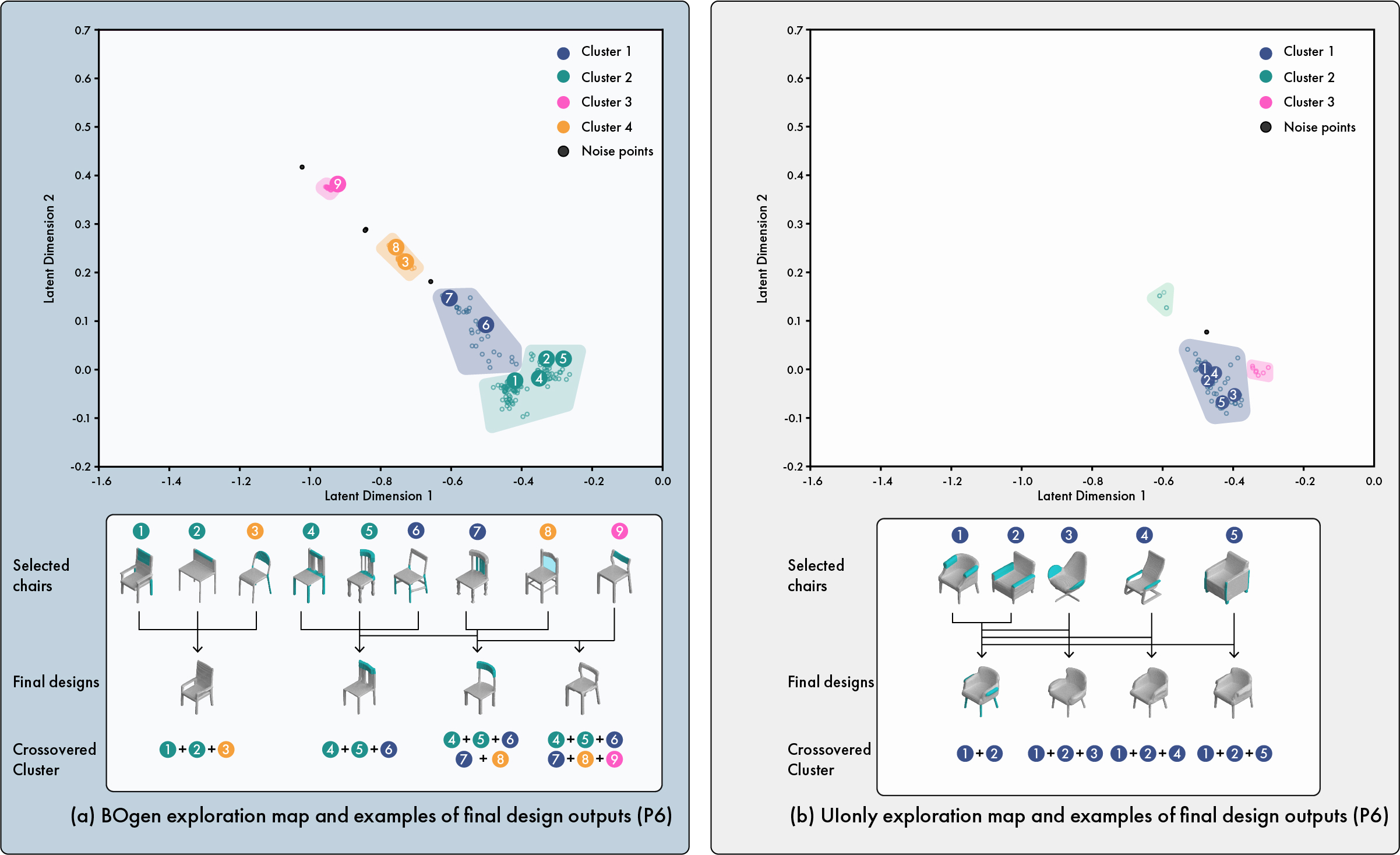}
  \caption{Exploration map and examples of final design outputs of P6: (a) BOgen; and (b) UIonly.}
  \label{fig:13}
\end{figure}

In the design context, as one iteratively explores diverse designs and converges on a specific design, the design process becomes novel \cite{goldschmidt2016linkographic}. In this regard, it is imperative to understand how densely a user has explored the design space and how diverse the exploration has been. We measured the area of the design space explored by users in each system and the number of clusters they explored. First, the exploration area in BOgen was significantly larger, approximately 120\% more extensive than that in UIonly (BOgen: 1.937, UIonly: 1.603, \( p<0.05 \); Figure \ref{fig:10}-c). Next, BOgen produced more clusters (approximately 127\%) than the UIonly system with our DBSCAN parameter settings (BOgen: 3.6, UIonly: 2.833, \( p<0.05 \); Figure \ref{fig:10}-d). In BOgen, users synthesize designs by crossing multiple clusters, notably. For example, in figure \ref{fig:13}-a, looking at P6’s use case with the BOgen system, P6 explored a total of four clusters. P6 developed designs by crossing over clusters 1 and 3 and clusters 1 and 2 (\ding{172}+\ding{173}+\ding{174}+\ding{175}+\ding{176}+\ding{177}). Importantly, designs developed from cluster 1, 2, and 3 were synthesized with cluster 4 before ending the experiment (\ding{174}+\ding{175}+\ding{176} +\ding{177} +\ding{178} +\ding{179}). In contrast, in \ref{fig:13}-b for the UIonly system, designs were synthesized only within cluster 1, with designs centered around the synthesis of \ding{172} and \ding{173}, limited to a narrow area. P6’s interview reflected: ’(In BOgen) I searched in broad categories first, then worked in detail based on the map. Some of the design options were synthesized to refine the design idea. With the map, I felt that I explored a more diverse design space than with the other systems (UIonly). Without the map (UIonly), it felt limiting and repetitive." The results indicated that through BOgen, users explored a broader, denser, and more varied design space. These advantages of BOgen were also evident in in-depth interviews with other participants:

\begin{itemize}
    \item P7: \textit{"In the system (UIonly), I could only see a limited range of chairs that appeared in the search, so I wasn’t sure about the designs available. However, in the other system (BOgen), through the map and its colors, I could check areas of interest and even areas that I did not care about, allowing me to see the overall designs."}
    \item P12: \textit{"In the system (BOgen), I could quickly and easily check what I wanted within a set of similar designs. The designs I had seen previously were marked on the map, which helped me think about exploring similar designs."}
    \item P27: \textit{"The addition of a map allowed me to check distributed points to see chair features. When ideas were limited to searches alone, new ideas were discovered in less interesting areas (lighter parts) on the map. In addition, since similar designs were closely distributed, I could find the design I wanted without searching multiple times and verify new design ideas."}
\end{itemize}
To summarize, BOgen not only guides the navigation of the design space for refining design ideas but also assists designers in finding prominent design parts. Additionally, it enables designers to delve into a dense and varied design space, thereby ensuring that users are more proactive in their design exploration.

\subsubsection{Implications}
Recent advancements in 3D generative artificial intelligence (AI) have empowered designers and general users to easily explore high-quality design variations. However, compared to 2D generative AI, 3D generative AI has complex and challenging characteristics \cite{liu20233dall,sanghi2023clip}. Therefore, the generation and synthesis of part-level designs that satisfy user requirements have significant implications. Fundamentally, design is a combination of elements
through which designers assess their appropriateness and creativity. In this context, our experimental results offer crucial insights into both the design process and design industry. 

Initially, interfaces supporting iterative design processes for 2D and 3D generative AI were not sufficiently developed. To tackle this, we created a UIonly interface to support the evolution of designers' thought processes. Building on this, we enhanced the system with BOgen, implementing an information model that offers recommendations based on user actions. However, under the UIonly condition, users were limited to generating
and synthesizing within their knowledge range, leading to a superficial exploration of the design. Conversely, BOgen enables designers to explore denser and more diverse design spaces, creating designs with prominent design elements based on part-level recommendations. Particularly in the UIonly situation, designers perceived the design space as only what the generative AI could create based on prompts, potentially missing unexplored design possibilities and limiting the discovery of desired part elements. However, BOgen’s exploration map enables designers to understand the entire design space of generative AI, visualize their design preferences, and accurately recommend part-level designs, thereby fostering a more proactive design process.

Furthermore, 3D design contains significantly more information than 2D design, providing designers with deeper insights. Specifically, 3D design objects enable simulations such as structural and topology simulations, production cost, and aerodynamic evaluations, which are limited to 2D designs. Thus, BOgen has the potential to innovate in the design industry. It effectively identifies preferred areas in a complex 3D design space based on designer preferences and generates suitable part-level 3D design outputs. Although modeling a desired 3D shape is a challenging task, even for experts, BOgen simplifies this complex and difficult process using 3D generative AI. By reflecting designer preferences, BOgen creates feasible and plausible 3D design outcomes by exploring previously unconsidered design combinations. This is directly linked to design prototyping and digital fabrication processes, enabling general users to effectively implement and produce user-customized 3D designs. Our findings will allow designers and the digital fabrication industry to more easily access and engage in 3D design, thereby significantly contributing to the design industry.

\section{CONCLUSIONS}
In this study, we introduced the UIonly and BOgen systems designed to assist in the development of
designers’ processes using 3D generative AI. We emphasized BOgen, empowering designers to actively generate
and explore part-level 3D designs. It utilizes preferential Bayesian optimization and a variational autoencoder,
incorporating user behavior and an exploration map to enhance the design experience and enable efficient
generation and synthesis. We effectively reduced the 3D generative AI latent space to a navigable 2D latent space
and proposed a method to create an exploration map for efficiently inferring users’ desired designs. The exploration
map and PBO-based recommendation features of BOgen showed three main findings in comparison with UIonly
from the experimental results. First, BOgen supported design-space navigation for the development of design ideas
by significantly predicting the uncertainty of the information space and probability of the selected design. Second,
BOgen successfully recommends the design parts desired by the designer by supporting sequential design
exploration, reflecting the changing intentions of the designer. Finally, BOgen supports designers in exploring and
synthesizing denser and more diverse designs. By exploring the impact of 3D Generative AI on the design process,
this study has significant implications for the design industry. BOgen enables the generation of customized designs
and part-level recommendations in a complex 3D design space, thereby reflecting designer preferences. This
approach allows a broader and more diverse exploration of the design space compared to the traditional UIonly
approach, offering creative design combinations that were previously unexplored. Such methods effectively assist
designers in evolving their design ideas and simplifying complex 3D design tasks that are limited to 2D designs.
In the broader context of the design industry, BOgen contributes to reducing the costs of design prototyping and
digital fabrication processes, while realizing user-customized designs. This represents a significant shift in the
design industry, improving the efficiency and creativity of 3D design processes for both designers and users.
Finally, BOgen allows users to explore designs more proactively. In the future, by integrating various modalities
including prompting, sketching, and eye-gaze-based interactions, we can better understand user intentions and
support a more designer-centric process in 3D generative AI.

\section{Acknowledgement}
This  work  was  supported  by  National  Research  Foundation  of  Korea  (NRF)  grant  funded  by  the  Korean government (MSIP: Ministry of Science, ICT and Future Planning) (RS-2023-00208542).

\bibliographystyle{unsrt}  
\bibliography{references}

\end{document}